\documentclass[sigconf]{acmart}

\usepackage{booktabs} 

\usepackage{verbatim}
\usepackage{graphicx}
\usepackage{subcaption}
\usepackage{url}
\usepackage{balance}
\usepackage{soul} 

\copyrightyear{2018} 
\acmYear{2018} 
\setcopyright{acmlicensed}
\acmConference[JCDL '18]{The 18th ACM/IEEE Joint Conference on Digital Libraries}{June 3--7, 2018}{Fort Worth, TX, USA}
\acmPrice{15.00}
\acmDOI{10.1145/3197026.3197040}
\acmISBN{978-1-4503-5178-2/18/06}

\begin{document}
\title{Extracting Scientific Figures with\\Distantly Supervised Neural Networks}


\author{Noah Siegel}
\affiliation{%
  \institution{Allen Institute for Artificial Intelligence}
  \streetaddress{}
  \city{Seattle}
  \state{Washington}
 }
\email{siegeln@uw.edu}
\authornote{Now at DeepMind.}

\author{Nicholas Lourie}
\affiliation{%
  \institution{Allen Institute for Artificial Intelligence}
  \streetaddress{}
  \city{Seattle}
  \state{Washington}
 }
\email{nicholasl@allenai.org}

\author{Russell Power}
\affiliation{%
  \institution{Allen Institute for Artificial Intelligence}
  \streetaddress{}
  \city{Seattle}
  \state{Washington}
 }
\email{russell.power@gmail.com}

\author{Waleed Ammar}
\affiliation{%
  \institution{Allen Institute for Artificial Intelligence}
  \streetaddress{}
  \city{Seattle}
  \state{Washington}
 }
\email{waleeda@allenai.org}

\begin{abstract}

Non-textual components such as charts, diagrams and tables provide key information in many scientific documents, but the lack of large labeled datasets has impeded the development of data-driven methods for scientific figure extraction.
In this paper, we induce high-quality training labels for the task of figure extraction in a large number of scientific documents, with no human intervention. 
To accomplish this we leverage the auxiliary data provided in two large web collections of scientific documents (arXiv and PubMed) to locate figures and their associated captions in the rasterized PDF.
We share the resulting dataset of over 5.5 million induced labels---4,000 times larger than the previous largest figure extraction dataset---with an average precision of 96.8\%, to enable the development of modern data-driven methods for this task.
We use this dataset to train a deep neural network for end-to-end figure detection, yielding a model that can be more easily extended to new domains compared to previous work.
The model was successfully deployed in {Semantic Scholar},\footnote{\url{https://www.semanticscholar.org/}} a large-scale academic search engine, and used to extract figures in 13 million scientific documents.\footnote{A demo of our system is available at \url{http://labs.semanticscholar.org/deepfigures/}, and our dataset of induced labels can be downloaded at \url{https://s3-us-west-2.amazonaws.com/ai2-s2-research-public/deepfigures/jcdl-deepfigures-labels.tar.gz}. Code to run our system locally can be found at \url{https://github.com/allenai/deepfigures-open}.}

\end{abstract}

%





\maketitle

\section{Introduction}

Non-textual components (e.g., charts, diagrams and tables) provide key information in many scientific documents.
Previous research has studied the utility of figures in scientific search and information extraction systems; however, the vast majority of published research papers are only available in PDF format, making figure extraction a challenging first step before any downstream application involving figures or other graphical elements may be tackled.
While some venues (e.g., PubMed) provide figures used in recently published documents, it remains a problem for older papers as well as many other venues which only publish PDF files of research papers.




Recent years have seen the emergence of a body of work focusing on use cases for extracted figures (see Section \ref{related_work}). All of these downstream tasks rely upon accurate figure extraction. Unfortunately, the lack of large-scale labeled datasets has hindered the application of modern data-driven techniques to figure and table extraction.\footnote{For brevity, we use \emph{figure extraction} to refer to the extraction of both figures and tables in the remainder of the paper.}
Previous work on this task used rule-based methods to address the problem in limited domains.
In particular, \cite{pdffigures} extract figures in research papers at NIPS, ICML and AAAI, and \cite{pdffigures2} extend their work to address papers in computer science more generally; however, stylistic conventions vary widely across academic fields, and since previous methods relied primarily on hand-designed features from computer science papers, they do not generalize well to other scientific domains, as we show in section \ref{results}.

Our main contribution in this paper is to propose a novel method for inducing high-quality labels for figure extraction in a large number of scientific documents, with no human intervention.
Our technique utilizes auxiliary data provided in two large web collections of scientific documents (arXiv and PubMed) to locate each figure and its associated caption in the rendered PDFs.
The resulting dataset consists of 5.5 million induced labels with an average precision of 96.8\%.
The size of this dataset is three orders of magnitude larger than human-labeled datasets available for figure extraction in scientific documents.

To demonstrate the value of this dataset, we introduce \emph{DeepFigures}, a deep neural model for detecting figures in PDF documents, built on a standard neural network architecture for modeling real world images, ResNet-101.
Comparison to prior rule-based techniques reveals better generalization across different scientific domains. 
Additionally, we discuss a production system for figure extraction built on DeepFigures that is currently deployed in a large-scale academic search engine (Semantic Scholar)\footnote{\url{https://www.semanticscholar.org}} which covers multiple domains, illustrating the practical utility of our proposed approach.

Our main contributions are:
\begin{itemize}
\item We propose a novel method for inducing high-quality labels for figure extraction in large web collections of scientific documents.
\item We introduce, to the best of our knowledge, the first statistical model for figure extraction, using a neural network trained exclusively on our dataset with no human labels.
\item We release our figure extraction data, tool, and code for generating the datasets and extracting figures locally to facilitate future research in graphical information understanding in scientific documents.
\end{itemize}

\section{Related Work} \label{related_work}
In this section, we discuss two lines of related work in the literature. 
The first line focuses on extraction and understanding of figures in scientific documents, which motivate this work.
The second line reviews related neural models which we build on.
\subsection{Scientific Figures}

Recent years have seen the emergence of a body of work focusing on applications involving  figures in scholarly research papers. 
Researchers have considered a range of tasks from extracting the underlying data from plots \cite{choudhury2015line,figureseer}, to the use of figures in search engines and broader information extraction systems \cite{Choudhury2013AFS,Tsutsui2016AnalyzingFO,Wu2015PDFMEFAM}.

Figures are also a common topic of interest in medical domains, where they often contain graphical images such as radiology imaging. In \cite{Wu2015PDFMEFAM}, researchers introduced the system PDFMEF, which incorporated figures into its extracted information. In \cite{Tsutsui2016AnalyzingFO}, researchers classified figures of brain images in order to provide more relevant information to doctors studying Alzheimer's Disease. In \cite{Choudhury2013AFS}, researchers present a search engine for figures in chemistry journals, allowing scientists to more easily locate information that may not be expressed in text.
Shared tasks such as ImageCLEF \cite{Herrera2015OverviewOT,Herrera2016OverviewOT} also helped drive more attention to compound figure detection \cite{Yu2017AssemblingDN}, compound figure separation \cite{Tsutsui2017ADD}, medical image annotation \cite{kumar2016adapting}, among other tasks related to medical images.

All of these downstream tasks rely upon accurate figure extraction. Previous work on this task has focused only on limited domains; \cite{pdffigures} focused only on 3 AI conferences, and \cite{pdffigures2} concentrated on papers only within computer science. Stylistic conventions vary widely across academic fields, and as we show in section \ref{results}, features hand-designed for computer science papers do not generalize well across other scientific domains.

Our method for inducing figure extraction labels can be viewed as an application of distant supervision, a popular approach for generating noisy labels in natural language processing tasks.
The key idea is to project known labels from an existing resource to related, unlabeled instances.
For example, \cite{Mintz2009DistantSF} used relations between entities in Freebase \cite{Bollacker2008FreebaseAC} to induce labels between pairs of corresponding entity mentions in unlabeled text, making the strong assumption that the Freebase relation is described in each sentence where both entities are mentioned.
As will be discussed in the following section, the method we propose does not require making such strong assumptions.

\subsection{Neural Models for Computer Vision}

The model architecture we use for figure extraction in this paper leverages the great success of convolutional neural networks on a variety of computer vision tasks including object recognition and detection \cite{resnet}, motion analysis \cite{Dosovitskiy2015FlowNet}, and scene reconstruction \cite{Yang2016Realtime3S}.
Inspired by the brain's visual cortex, these networks consist of millions of neurons arranged in a series of layers that learn successively higher-level visual representations. 
For example, when performing facial recognition, a neuron in the first layer might detect horizontal edges, a neuron in the second layer might respond to certain curves, the third layer to an eye, the fourth layer to a whole face. Recent work has used neural networks for semantic page segmentation, suggesting that these models can be applied to synthetic documents as well as natural scenes \cite{Chen2015ICDAR,Yang2017CVPR,He2017ICDAR}.
In this section we provide more details on two building blocks we use in DeepFigures: ResNet-101 and OverFeat.

\subsubsection{ResNet-101}

An important task in computer vision is object recognition: for example, given an image, determine whether it is a cat or a dog. Using the raw pixels as features poses difficulties for most traditional classification algorithms, due to the sheer volume of information and the curse of dimensionality. Instead, computer vision techniques generally extract higher level features from the image, and then run a standard machine learning classifier such as logistic regression on these features. Before neural networks, features were generally hand-engineered by researchers or practicioners; an example of one common such feature is the frequencies of edges in various regions of the image \cite{Dalal2005HOG}. In contrast, convolutional neural networks learn their feature representations from the data. This learning is achieved by defining a broad space of possible feature extractors and then optimizing over it, typically using backpropagation and stochastic gradient descent. The architecture of the neural network corresponds to how the neurons are defined and pass information to each other and it is the neural network's architecture that defines the space of possible feature extractors we might learn.

Numerous highly successful neural network architectures have been proposed for computer vision \cite{Krizhevsky2012ImageNetCW,Szegedy2015,resnet}. Generally, with more data and more layers, neural networks tend to get increased performance. Because of this fact, large-scale datasets and optimization methods are key to neural networks' success. One problem in training neural networks with many layers is that of vanishing gradients: as gradients are propagated through successive layers, they tend to either blow up (causing parameters to quickly diverge to infinity during training) or shrink to zero, making training earlier layers difficult. Residual networks (ResNets) \cite{resnet} address the problem by adding identity connections between blocks: rather than each layer receiving as input only the previous layer's output, some layers also receive the output of several layers before. 
These identity connections provide a path for gradients to reach earlier layers in the network undiminished, allowing much deeper networks to be trained. 
An ensemble of ResNet models won the ImageNet object detection competition in 2015. 

Because useful image features transfer well across tasks it is common to use parts of one neural network architecture in place of components of another. 
ResNet-101 \cite{resnet} provides one such feature extraction architecture. ResNet-101 is a 101-layer deep neural network formed by stacking ``bottleneck'' units consisting of a 1x1 convolutional layer, followed by a 3x3 convolutional layer that brings down the dimension of the embedding, and then another 1x1 convolutional layer that brings the dimension of the embedding back up to that of the original input. An identity connection adds the input of the bottleneck unit into the output of its last layer before passing it further down the network.

\subsubsection{OverFeat} Another common task in computer vision, and the one in which we are interested in this work, is object detection: for example, determine the location of all human faces in a given image. 
This task can be formalized as predicting a bounding box that encloses the object while being as small as possible. 
Object detection is more complex than classification since rather than predicting a binary or multiclass label, the model must predict a variable-length list of bounding box coordinates that may change in size and shape. The problem can be reduced to classification by running a classifier on every possible box on the image, but due to the high computational cost of running neural networks with millions of parameters this is generally infeasible.

OverFeat \cite{overfeat} introduced the idea of bounding box regression. 
Rather than producing a class output, the model can use regression to predict bounding box coordinates directly. To enable detecting multiple objects as well as to handle objects in various locations in the image, the model is run fully-convolutionally, i.e., the entire model is run on cropped image sections centered on a uniformly spaced grid of 20x15 points.\footnote{Running the model at each point on a grid is significantly less computationally expensive than running the model on the same number of independent images, because the convolutional structure of network layers means much of the work on overlapping regions is redundant and can be shared; see \cite{overfeat} for details.}
For each cropped region, in order to extract the feature vectors, OverFeat uses 5 initial layers that perform convolutions and max pooling. Classification is then performed by two fully connected layers and an output layer from the feature vectors; while bounding box regression is performed by two fully connected layers and a final output layer providing 4 numbers -- the coordinates for the predicted bounding box. Each class has its own output layer to provide a bounding box for that class alone. The classification result then provides a confidence for each class and every region in the grid, while the bounding box regression yields a possible bounding box for that class. Thus, for any class many bounding boxes are predicted and then later merged.

\begin{figure*}
\includegraphics[width=.33\linewidth]{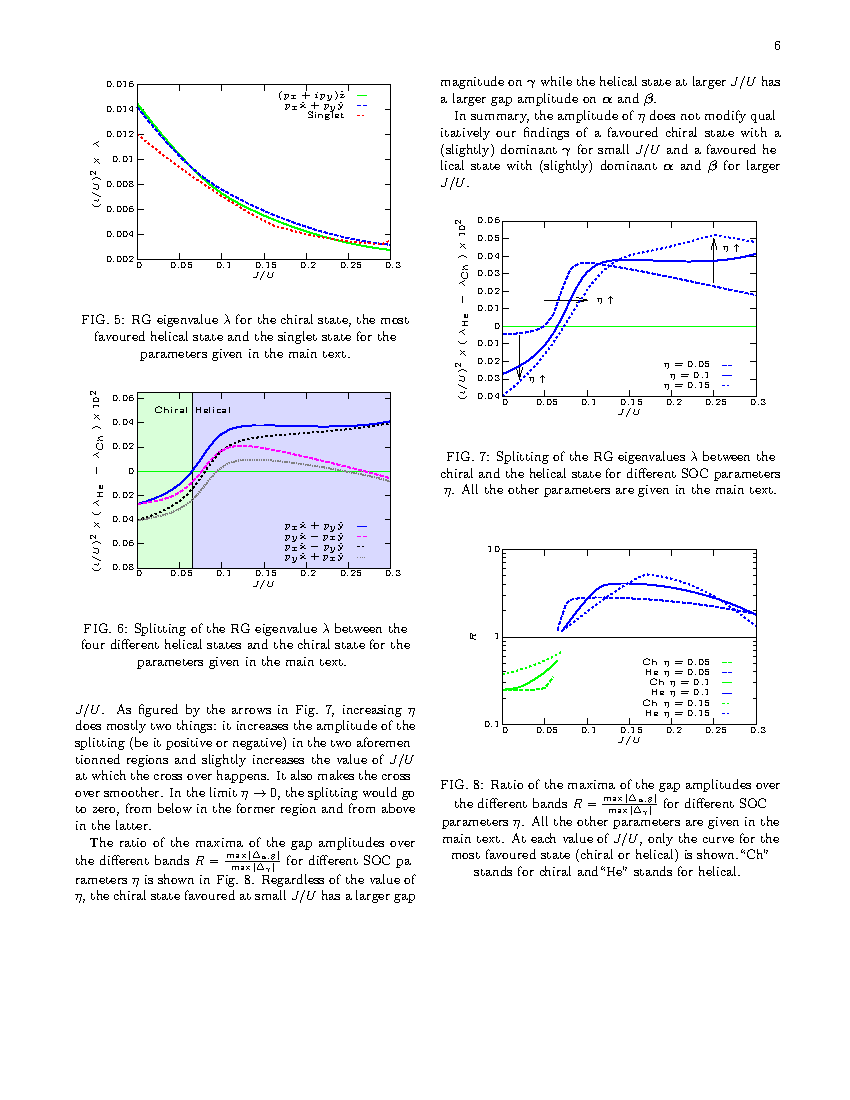}
\includegraphics[width=.33\linewidth]{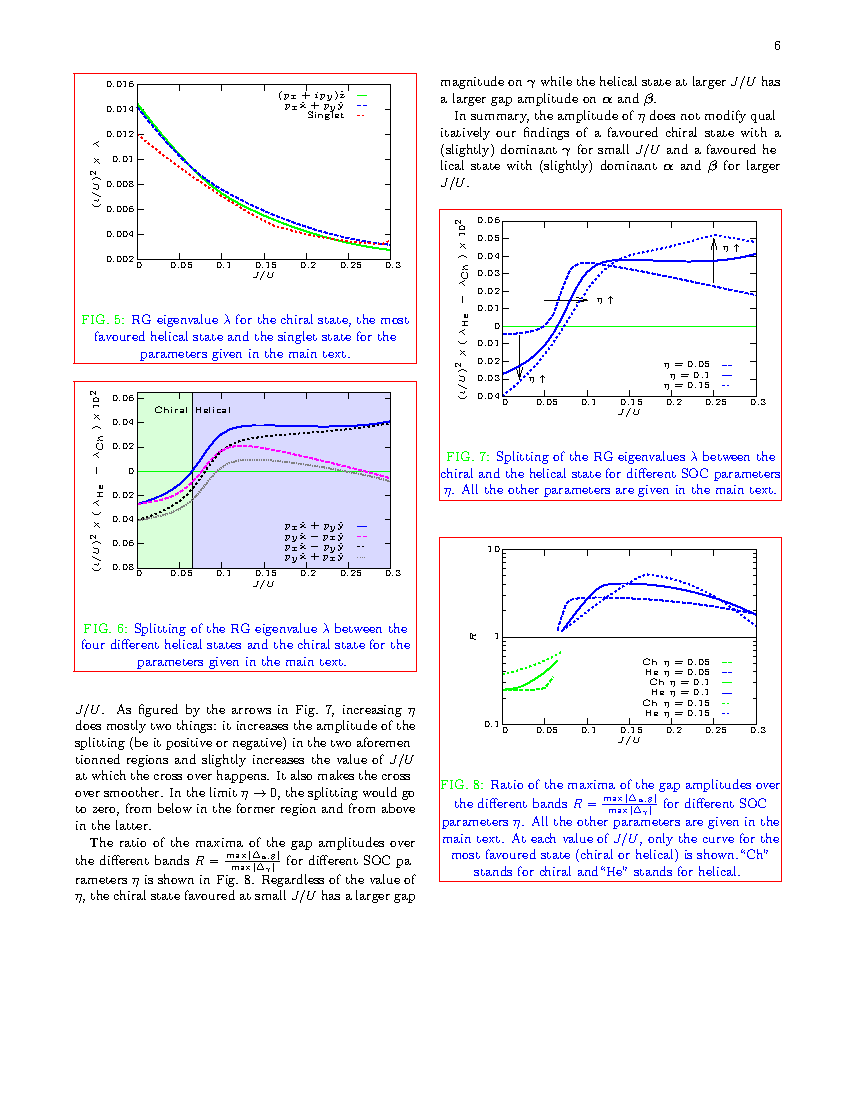}
\includegraphics[width=.33\linewidth]{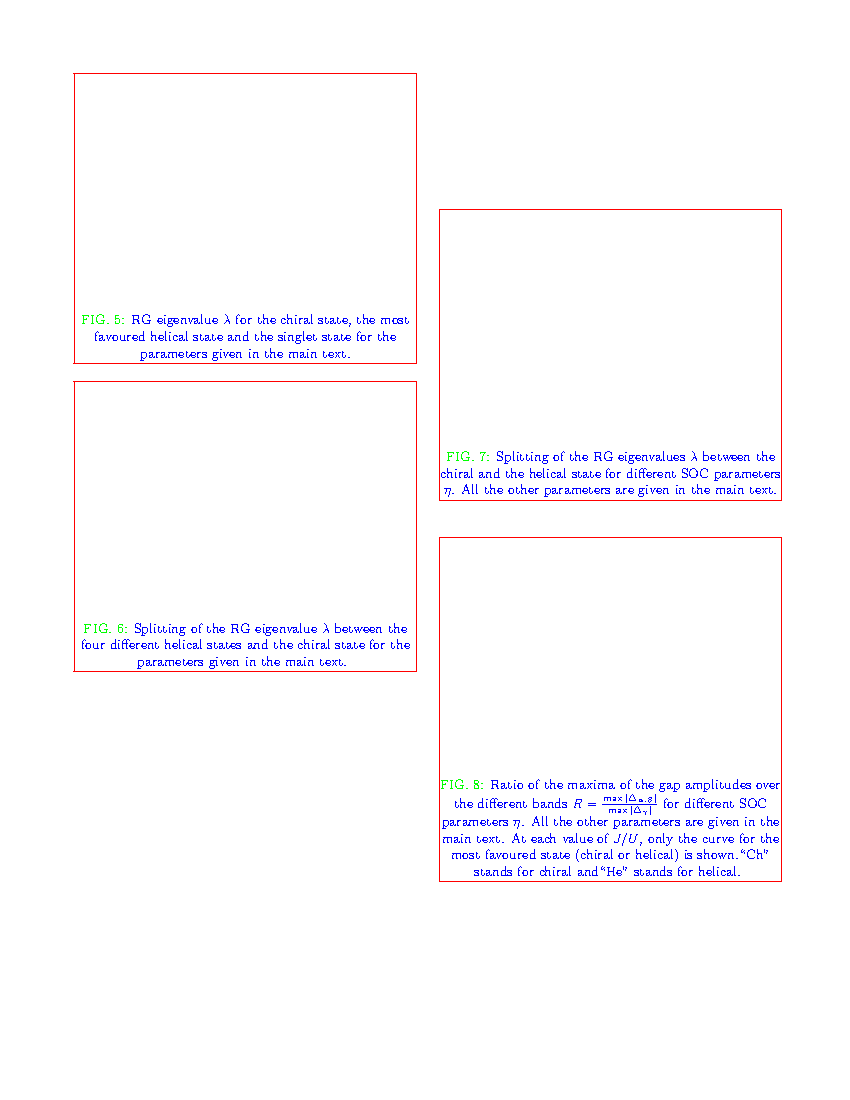}
\caption{Modifying LaTeX source to recover figure positions. Figure bounding boxes are shown in red, figure names in green, and captions in blue. Left: original document. Middle: document compiled from modified source. Right: image difference between original and modified documents.}
\label{fig:latex}
\end{figure*}

\section{Inducing Figure Extraction Labels}\label{data}

A key contribution of this paper is a novel method for inducing high-quality labels for figure extraction in a large number of scientific documents (see Table \ref{tab:dataset_stats} for dataset statistics).
The resulting dataset is critical for training statistical models for figure extraction, especially  for deep neural networks, e.g., \cite{resnet}.
In order to induce the labels, we align the figures and tables specified using a markup languages (e.g., LaTeX) with bounding boxes in the corresponding PDF files, then use coordinates of the bounding boxes as labeled data.
The following subsections provide details on how to do this alignment using two markup languages commonly used in the scientific literature: LaTeX and XML.

\subsection{Aligning Figures in LaTeX Documents}
\label{sec:align_latex}
Many researchers use LaTeX to typeset scientific documents, before compiling them into PDFs.\footnote{We use the term \emph{LaTeX} to refer to the formal language used to describe a document's content, structure and format in \TeX~software distributions.}
In LaTeX, figures and tables are specified using standard commands such as \verb|\includegraphics|.
We modify the way figures render in the compiled PDF file by adding special instructions in the header section of LaTeX source files which results in drawing a rectangle around each figure and table in the rendered PDF file. 
We then use the pixel-by-pixel image difference between the original and modified versions of the rendered PDF file to identify the bounding boxes of each figure, as illustrated in Fig.~\ref{fig:latex}.

Unlike plain text, which may be separated by line and page breaks at the compiler's discretion, figures and tables need to be represented in contiguous blocks of space. 
To handle these graphical elements, LaTeX uses the concept of ``floats''. 
Floats are not part of the normal stream of text, but are instead placed at a location determined by the LaTeX compiler.
Floats also include a caption to describe them and a name (e.g. ``Figure 1.''), allowing them to be referenced from elsewhere in body text. 
\\[15pt]
\textbf{Labeling figures and tables:}

We add the following commands to the header section of the LaTeX source file to surround each figure and table with a bounding box in the compiled PDF file:

\begin{verbatim}

\usepackage{color}
\usepackage{floatrow}
\usepackage{tcolorbox}

\DeclareColorBox{figurecolorbox}{\fcolorbox{red}{white}}
\DeclareColorBox{tablecolorbox}{\fcolorbox{yellow}{white}}

\floatsetup[figure]{framestyle=colorbox,
    colorframeset=figurecolorbox, framearound=all,
    frameset={\fboxrule1pt\fboxsep0pt}}
\floatsetup[table]{framestyle=colorbox,
    colorframeset=tablecolorbox, framearound=all,
    frameset={\fboxrule1pt\fboxsep0pt}}
\end{verbatim}

We use different colors for figures and tables, in order to differentiate these types automatically.
Recompiling and computing the image difference between the modified and the original PDF files yield blank pages containing only the boxes.\footnote{The added border shifts figure positions by a few pixels, so we use the same command to add white borders to figures in the original to make them align exactly.} We can then locate boxes by finding all connected regions of non-blank pixels and then taking the minimum and maximum x and y coordinates of each.
\\[15pt]
\textbf{Labeling captions:}

In order to find the coordinates of the bounding box for caption text, we modify the color of figure names and captions using the following command:
\begin{verbatim}
\usepackage[labelfont={color=green},
    textfont={color=blue}]{caption} 
\end{verbatim}

Finally, we modify the coordinates of the bounding box of each figure and table to exclude the caption, by identifying the largest rectangular region inside the float that contains no caption text.
This is robust even to uncommon caption locations (e.g., above a table, or to the side of a figure).
\\[15pt]
\textbf{arXiv:}

In order to construct our dataset, we download LaTeX source files from arXiv,\footnote{\url{https://arxiv.org/}} a popular platform for pre-publishing research in various fields including physics, computer science, and quantitative biology.
When authors make a paper submission to arXiv, they are required to upload source files if the paper is typeset using LaTeX.
As of the time of writing, arXiv hosts over 900,000 papers with LaTeX source code.

\subsection{Aligning Figures in XML Documents}
We would like our dataset to cover a diverse set of scientific domains.
Although LaTeX is widely used in some fields (e.g., statistics, physics, computer science, etc.), it has been less popular in important domains such as the medical and life sciences where WYSIWYG editors (e.g., Microsoft Word) are more common. 
As a result, we cannot use the method described in section \ref{sec:align_latex} to induce a large labeled dataset in these fields.
\\[15pt]
\textbf{PubMed:}

Fortunately, however, some publishers provide XML markup for their papers, which can also be used to induce figure extraction labels. 
In particular, PubMed Central Open Access Subset is a free archive of medical and life sciences research papers. 
The National Center for Biotechnology Information (NCBI) makes this subset available through bulk downloading.
In addition to the PDF files, it provides auxiliary data to improve the user experience while reading a paper. 
The auxiliary data includes the paper text marked up with XML tags (including figure captions) as well as image files for all graphics. 

In principle, this data can be used to induce labels for figure extraction.
However, unlike LaTeX documents, the XML markup cannot be used to compile the PDF.
Therefore, we propose a different approach to recover the positional information of figures.
\\[15pt]
\textbf{Labeling captions:}

First, for each image in the auxiliary data, we determine which page in the corresponding PDF file contains this figure by searching the PDF text (extracted by a tool such as PDFBox) for the caption text (which is also available in the auxiliary data).
Since the XML and PDF text do not always match exactly (e.g., em dash in PDF vs. a hyphen in XML), we use dynamic programming to find the substring in the PDF text with smallest Levenshtein distance to the caption text in the XML file.
We modify the standard Wagner-Fischer dynamic programming algorithm for edit distance \cite{wagner1974string} by setting the cost for starting (and ending) at any position in the PDF text to 0.
This modification maintains the time complexity of $O(mn)$, where $m$ and $n$ are the string lengths. 
\\[15pt]
\textbf{Labeling figures:}

Once we have identified the page that a figure is on, we render that page as an image and then use multi-scale template matching \cite{brunelli2009template} to find the position of the figure on the page.
We use the figure image as a filter and cross-correlate it with the page to produce a spatial map of image correlations. 
Template matching is typically done using image representations such as edge detections or oriented gradients \cite{brunelli2009template}, but because we do not have to deal with typical conditions present in natural images such as variations in lighting or pose, we find that template matching in raw pixel space works best. 
We use OpenCV's \texttt{matchTemplate} implementation with the similarity metric \texttt{CV\_TM\_CCOEFF\_NORMED}, matching at 45 scales where the figure's largest dimension relative to the page takes up between 10\% and 95\% of the page. 

In rare cases, the provided figure images do not match the figures as they appear in the PDF (e.g., subfigures may be laid out horizontally on the PDF but vertically in the provided image file). 
If template matching yields a similarity below 0.8 for any figure, we exclude the paper from our dataset to reduce the risk of inaccurate training data.
\\[15pt]
\textbf{Labeling tables:}

Tables are sometimes provided as images in the same way figures are.
However, it is more common for tables to be represented directly in the XML with tags for each table cell. 
We first tried using the textual edit distance to identify table coordinates in the PDF file (similar to captions), but we found that the order of table cells often differs between PDFBox's extracted text and the XML (e.g. table cells may be extracted from the PDF in column-major order while the XML is row-major). 
Therefore, we instead use a bag of words similarity. 

We find the token sequence in the PDF that has the highest similarity to the set of words in the XML table. 
We can find the optimal sequence in the PDF text efficiently by maintaining a word difference counter. 
For a given start position in the PDF stream, we initialize the counter to the bag of words from the XML table.
For each token following this position, we decrement the counter for the word at the current position (while allowing negative counts to represent words that occur more in the PDF than in the XML). 
This procedure is repeated for each start position on the PDF page. The following pseudo-code illustrates our algorithm for finding the interval on the page with the lowest bag-of-words distance to the table:
\begin{verbatim}
best_dist <- math.inf
for start_word in page_words:
  diff_counter <- table_words
  cur_dist = sum(table_words)
  for end_word in page_words from start_word:
    diff_counter[end_word] -= 1
    if diff_counter[end_word] >= 0:
      cur_dist -= 1
    else:
      cur_dist += 1
    if cur_dist < best_dist:
      best_dist <- cur_dist
      store start_word and end_word positions
\end{verbatim}
If $m$ is the length of the XML table and $n$ is the length of the PDF, generating or copying the initial word counter is $O(m)$ and iterating over ending words on the PDF text is $O(n)$. Both of these occur for every starting word, for a total time complexity of $O(n(n+m))$. 
We identify the minimum axis-aligned bounding box containing all caption tokens as the table caption region.

\begin{figure*}
\includegraphics[width=.49\linewidth]{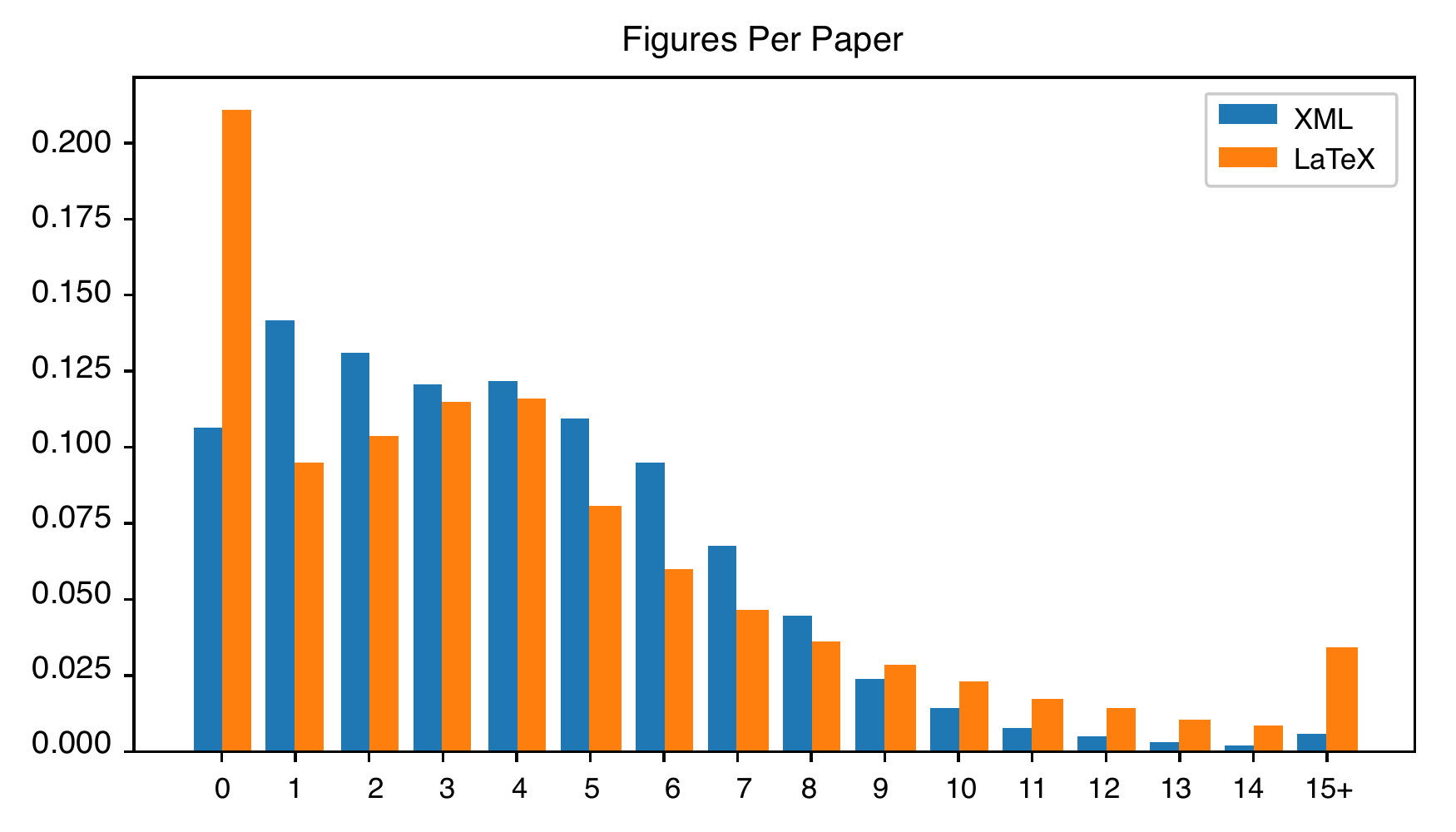}
\includegraphics[width=.49\linewidth]{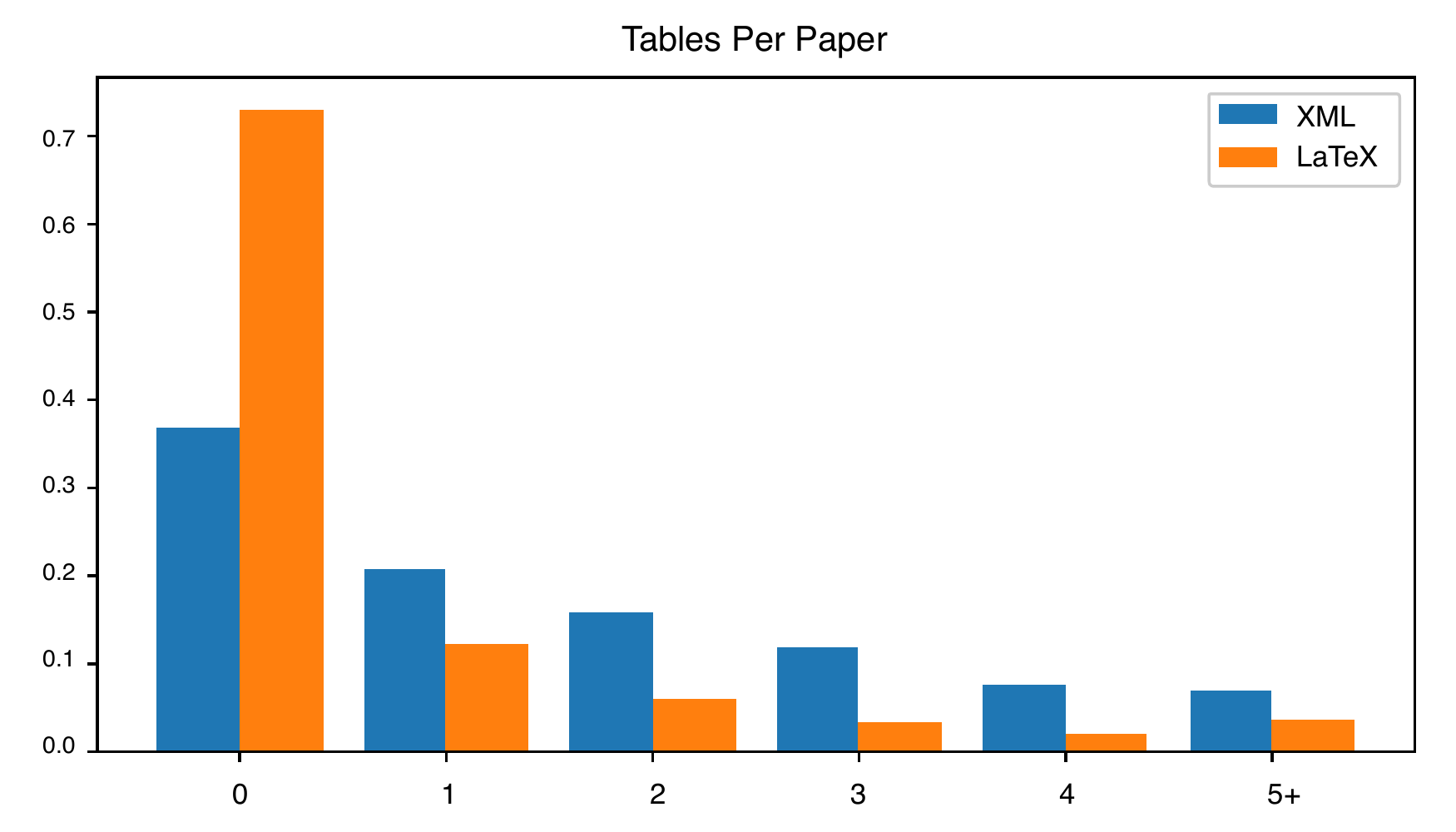}
\caption{Distributions of figures (left) and tables (right) in our automatically generated datasets. X-axis shows number of figures/tables per paper. Y-axis shows fraction of papers with that many figures/tables. Differences are likely a result of the differing source datasets: for example, the life science papers found in PubMed may rely more on tables to convey information than math papers on arXiv.
}
\label{fig:histogram}
\end{figure*}

\begin{table}[t]
\begin{center}
\begin{tabular}{|l|rr|rr|} \hline
Dataset        & \multicolumn{2}{c|}{Manually-labeled} & \multicolumn{2}{c|}{Induced labels} \\
name & CS-Large \cite{pdffigures2} & PubMed & LaTeX & XML \\ \hline 
\# papers & 346   & 104      & 242,041 & 791,381   \\ 
\# figures & 952   & 289      & 1,030,671     & 3,064,951   \\ 
\# tables  & 282   & 124      & 164,356     & 1,267,464  \\ \hline
\end{tabular}
\vspace{0.2cm}
\caption{Number of papers, figures, and tables in the manually-labeled datasets (left) and our datasets of induced labels (right).}\label{tab:dataset_stats}
\end{center}
\end{table}

\subsection{Comparison to Manual Annotation}
\label{sec:dataset_stats}
In this section, we proposed a method for automatically inducing labeled data for figure extraction in scientific documents.
An alternative approach is to train annotators to sift through a large number of research papers and label figure and table coordinates and their captions.
While this approach typically results in high quality annotations, it is often impractical.
Manual annotation is slow and expensive, and it is hard to find annotators with appropriate training or domain knowledge.
With limited time and budget, the size of labeled data we can collect with this approach is modest.\footnote{Another alternative is to use crowdsourced workers (e.g., using Amazon Mechanical Turk \url{https://www.mturk.com/} or CrowdFlower \url{http://www.crowdflower.com/}) to do the annotation.
Although crowdsourcing has been successfully used to construct useful image datasets such as ImageNet \cite{ImageNet}, \cite{figureseer} found that crowdsourcing figure annotations in research papers yielded low inter-annotator agreement and significant noise due to workers' lack of familiarity with scholarly documents.}
\\[15pt]
\textbf{Scalability of induced labels:}

In contrast to manual annotation, our proposed method for inducing labels is both scalable and accurate.
We compare the size of our datasets with induced labels to that of manually labeled datasets in Table \ref{tab:dataset_stats}.
We compare with two manually labeled datasets:
\begin{itemize}
\item The ``CS-Large'' dataset \cite{pdffigures2}: To our knowledge, this was previously the largest dataset for the task of figure extraction. Papers in this dataset were randomly sampled from computer science papers published after the year 1999 with nine citations or more.
\item The ``PubMed'' dataset: We collected this dataset by sampling papers from PubMed, and hired experts in biological sciences to annotate figures and tables, and their captions. 
\end{itemize}
Both manually labeled datasets are used as test sets in our experiments (section \ref{results}).
Notably, both of our datasets with induced labels ``LaTeX'' and ``XML'' are three orders of magnitude larger than ``CS-Large''.
\\[15pt]
\textbf{Accuracy of induced labels:}

In order to assess the accuracy of labels induced using our method, we collected human judgments for a sample of papers in the ``LaTeX'' and ``XML'' datasets.
Table \ref{tab:dataset_eval} reports the precision and recall of figures and tables, including captions, for 150 pages in 61 papers in the ``LaTeX'' dataset and 106 pages in 86 papers in the ``XML'' dataset.
We require that the four corners of a figure (or table) and the four quadrants of its caption must be correct for each true positive data point.
As shown in Table \ref{tab:dataset_eval}, the quality of induced labels are fairly high (e.g., the F1 score of induced labels ranges between 93.9\% and 100\%).

\begin{table}
\begin{center}
\begin{tabular}{|l|rr|rr|} \hline
Dataset        & \multicolumn{2}{c|}{LaTeX} & \multicolumn{2}{c|}{XML} \\
               & Figures & Tables & Figures & Tables \\ \hline 
Precision      & 1.00    & 1.00   & 0.97    & 0.94   \\ 
Recall         & 0.95    & 1.00   & 0.91    & 0.94   \\ 
F1             & 0.97    & 1.00   & 0.94    & 0.94   \\ \hline
\end{tabular}
\vspace{0.2cm}
\caption{Precision, recall and F1 score of induced labels in the ``LaTeX'' and ``XML'' datasets. }\label{tab:dataset_eval}
\end{center}
\end{table}

The following section discusses the model we developed in order to consume the induced labeled data described in this section.

\section{The DeepFigures Model} \label{sec:model}

\begin{figure*}
\includegraphics[width=.99\linewidth]{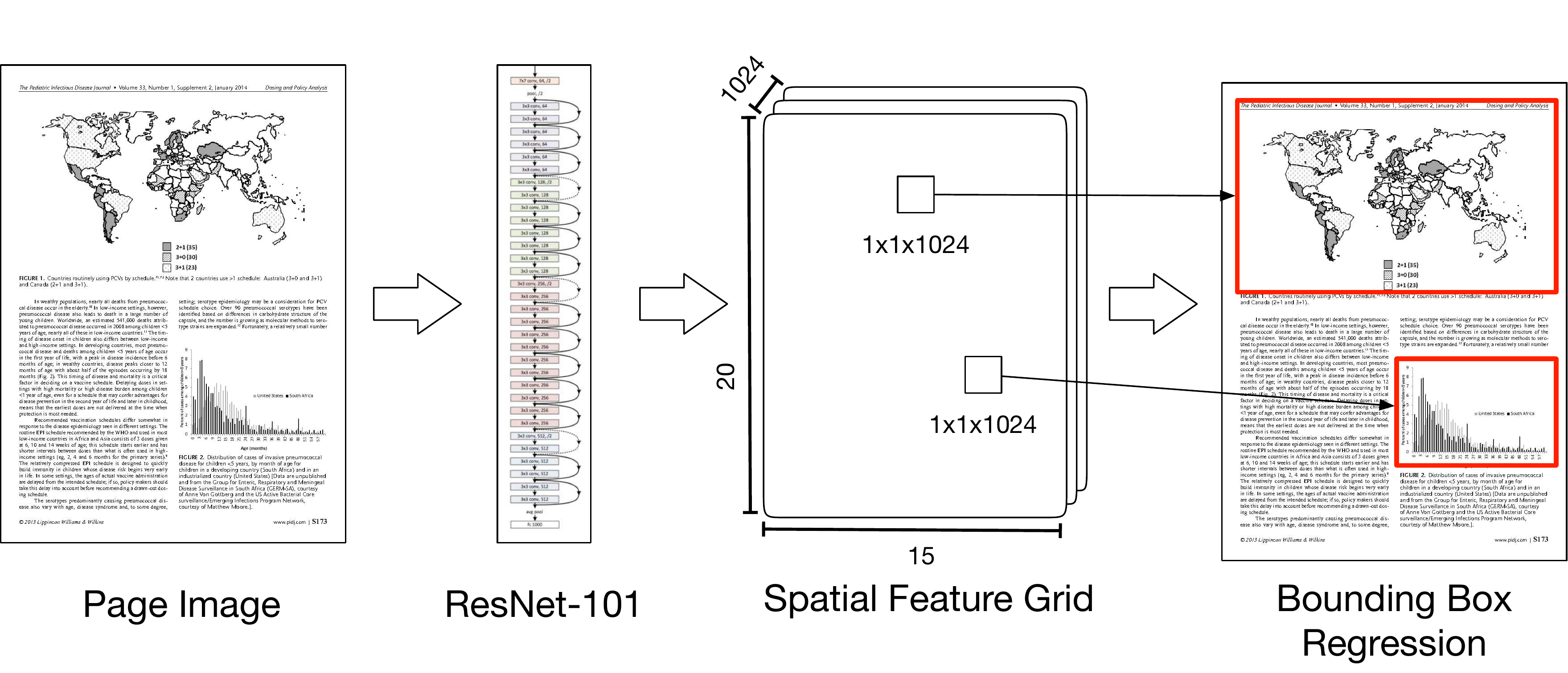}
\caption{High-level structure of the DeepFigures model. The input to the model is a 640x480 page image. ResNet-101 is run fully-convolutionally over the image, yielding a 20x15 spatial grid of 1024-dimensional image embedding vectors. Next, regressions to predict box coordinates and confidences are run on each of the 300 grid cells, yielding 300 candidate bounding boxes. Running non-maximum suppression and filtering out predictions with confidences below a threshold yields the final predictions.}
\label{fig:structure}
\end{figure*}

\begin{figure}
\includegraphics[width=.47\textwidth]{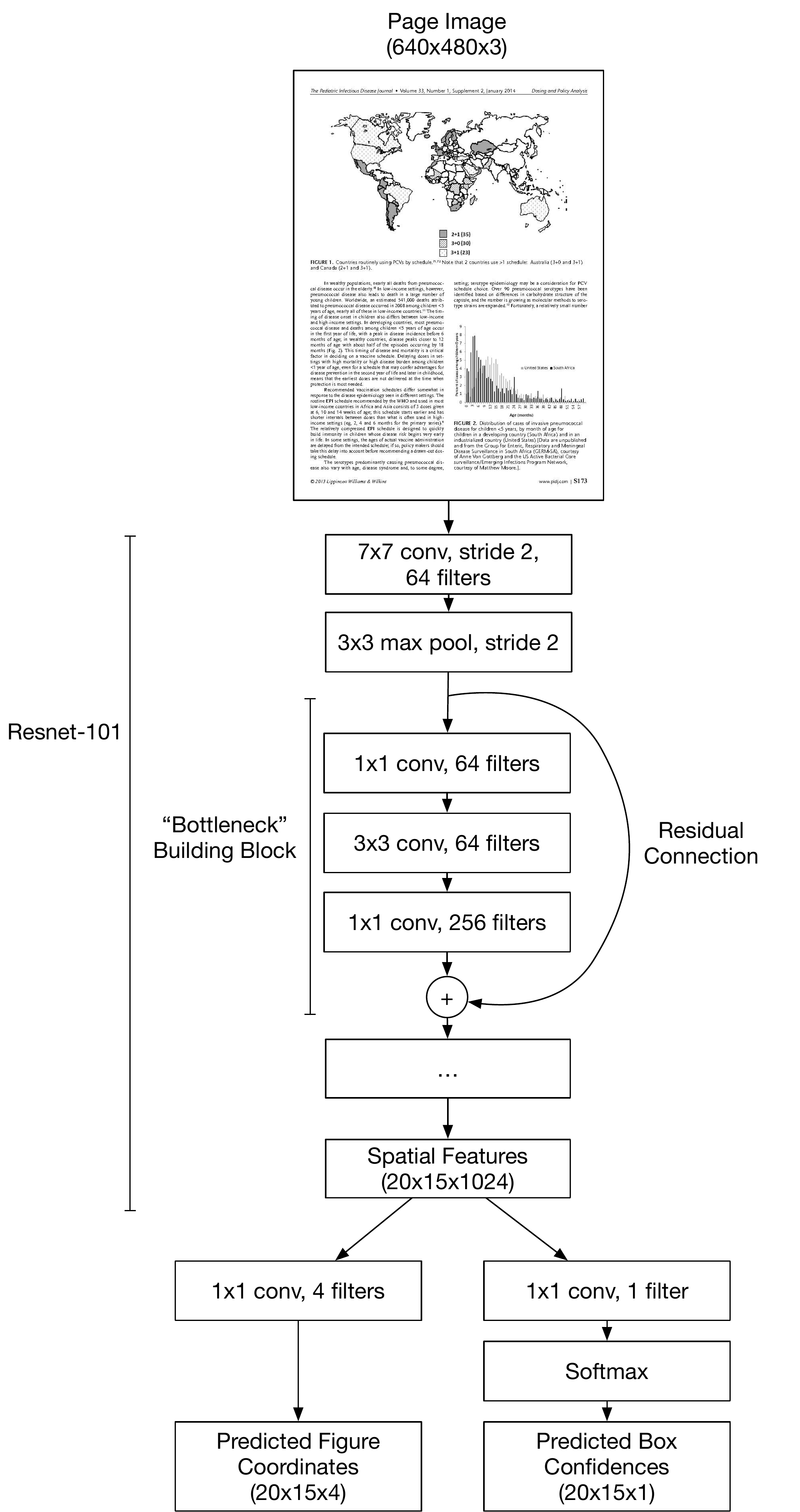}
\caption{Architecture of the DeepFigures network expressed fully convolutionally. Note that a 1x1 convolution is equivalent to a fully connected layer run at each point on a spatial grid. Strides are 1 where not specified and all convolutional layers except those outputting predictions use ReLU activations. See \cite{resnet} for the full ResNet-101 architecture.}
\label{fig:network}
\end{figure}

Our system takes as input a PDF file, which we then render as a list of page images, and feed each page to our figure detection neural network. 
The network architecture we use for figure extraction is a slight variant of several standard neural network architectures for image classification and object detection. In particular, our model is based on TensorBox \cite{tensorbox}, applying the OverFeat detection architecture \cite{overfeat} to image embeddings generated using ResNet-101 \cite{resnet}. This object detector then finds bounding boxes for figures in the PDF, and captions are extracted separately. 

In contrast to OverFeat, which uses a relatively shallow 5-layer network to generate the spatial feature grid, we use ResNet-101 which enables higher model capacity and accuracy. Additionally, while OverFeat trained the embedding network on a classification task and then fixed those weights while training localization, learning only the weights for the final regression layer, we train the full network end-to-end, allowing the embedding network to learn features more relevant to localization and eliminating the need for pre-training. 

As illustrated in Figure \ref{fig:structure}, in the network we use for figure extraction each grid cell is represented by 1024 features extracted from the ResNet-101 model, resulting in a 20x15x1024 spatial feature grid. 
At each grid cell, the model uses these features to predict both a bounding box and a confidence score. Boxes with confidence score above a selected threshold are returned as predictions. Figure \ref{fig:network} illustrates the architecture in more detail.
\\[15pt]
\textbf{Matching Captions:}

The OverFeat-ResNet figure detection model outputs a set of bounding boxes for figures; however, many applications, including academic search, benefit most from having a set of figure-caption pairs for the PDF. Our figure extraction pipeline extracts captions' text and bounding boxes using the same method as \cite{pdffigures2}, finding paragraphs starting with a string that matches a regular expression capturing variations of ``Figure N.'' or ``Table N.'' and then locating the identified textual elements in the page using standard PDF processing libraries. 
Once we have a list of proposed captions, we match figures to captions in order to minimize the total euclidean distance between the centers of paired boxes. 
This is an instance of the linear assignment problem and can be solved efficiently using the Hungarian algorithm \cite{kuhn1955hungarian}. 
If there are more detected figures than captions or vice-versa, the algorithm picks the $\min(\text{figure count, caption count})$ pairs that minimize total distance.
See \cite{pdffigures2} for more details on matching captions. 

Both of the data generation methods described in section \ref{data} produce bounding boxes for captions as well as figures, so in principle the captions could also be detected using a neural network. In our experience, however, training the model to predict captions reduced performance. There are a few likely causes: captions are often very small along the height dimension, amplifying small absolute errors in bounding box coordinates. Similarly, captions have fewer visual cues and are much less distinct from surrounding text than figures. Finally, the baseline caption detection model from \cite{pdffigures} performs very well. Most errors in PDFFigures 2.0 are caused by figure detection error rather than caption detection. For these reasons, we continue to use the rules-based approach for detecting captions. 

To summarize how our model combines the previously mentioned components, the model generates a 20x15 spatial grid of image embedding vectors with each embedding vector having 1024 dimensions generated using ResNet-101 \cite{resnet}.
The feature vectors are then input into a linear regression layer with two outputs that represent the four coordinates of the bounding box. Simultaneously, the feature vectors are passed through a logistic regression layer to predict the confidence that each grid cell is at the center of a figure. Redundant boxes are eliminated via non-maximum suppression. At test time, we run inference with a confidence threshold of 50\%, although this parameter may be tuned to favor precision or recall if needed.

\section{Experiments} \label{results}
In this section, we compare the DeepFigures model described in section \ref{sec:model} to PDFFigures 2.0 \cite{pdffigures2}, the previous state of the art for the task of figure extraction.
\\[15pt]
\textbf{Data:}

We train the DeepFigures model on 4,095,622 induced figures (1,030,671 in the LaTeX dataset and 3,064,951 in the XML dataset) and 1,431,820 induced tables (164,356 in the LaTeX dataset and 1,267,464 in the XML dataset).
See section \ref{data} for more details on the two datasets.

When using any algorithmically generated dataset, the question arises of how to ensure that the model is really learning something useful, rather than simply taking advantage of some algorithmic quirk of the generating process. Therefore, we perform evaluation entirely on human annotated figures. The algorithm is trained entirely on synthetic data and tested entirely on human labeled data, so our high performance demonstrates the quality of our distantly supervised dataset.

We run evaluation on two datasets: the ``CS-Large'' computer science dataset introduced by \cite{pdffigures2}, and a new dataset we introduce using papers randomly sampled from PubMed. Our new dataset, consisting of 289 figures and 124 tables  from 104 papers, was annotated by experts in biological sciences.
\\[15pt]
\textbf{Hyperparameters:}

We use RMSProp as our optimizer with initial learning rate 0.001. We train for 5,000,000 steps, decaying the learning rate by a factor of 2 every 330,000 steps. We use a batch size of 1 during training and did not observe significant performance gains from larger batches, likely due to the inherent parallelism in sliding window detection.
\\[15pt]
\textbf{Evaluation Methodology:}

Our evaluation methodology follows that of \cite{pdffigures2}. A predicted box is evaluated against a ground truth box based on Jaccard index, also known as intersection over union (IOU): the area of their intersection divided by the area of their union. As in \cite{pdffigures2}, a predicted figure bounding box is considered correct if its IOU with the true box exceeds .80, while a predicted caption box is considered correct if its IOU exceeds .80 or if the predicted caption text matches the text from the true region extracted from the PDF. However, while \cite{pdffigures2} required annotations to include the figure number in order to be matched with predictions, we eliminate this requirement in order to simplify the human annotation task. Instead, we find the optimal assignment of predicted figures to true figures for each page, which is an instance of the linear assignment problem and can be done efficiently using the Hungarian algorithm.
\\[15pt]
\textbf{Results:}

As shown in Table \ref{table:eval}, DeepFigures underperorms by 3 F1 points on ``CS-Large'', but achieves a 17 point improvement on the ``PubMed'' dataset. 
Given that PDFFigures 2.0 is a rule-based method that was tuned specifically for the ``CS-Large'' test set, it is unsurprising to see that it works better than DeepFigures for this domain. 

Since DeepFigures does not use any human annotation or domain specific feature engineering, it has learned a robust model for identifying figures across a variety of domains.
For example, PDFFigures 2.0 often generates false positives for graphical headers which are visually distinct from actual figures, however, allowing our model to correctly reject them.

\begin{table}
\begin{center}
\begin{tabular}{|l||r|r|} \hline
 System & CS-Large & PubMed\\ \hline 
 PDFFigures 2.0 \cite{pdffigures2} & \bf{87.9\%} & 63.5\%  \\ 
 DeepFigures (Ours) & 84.9\% & \bf{80.6\%} \\ \hline
\end{tabular}
\vspace{0.2cm}
\caption{F1-scores for figure extraction systems on human labeled datasets. In keeping with \cite{pdffigures2}, a predicted bounding box is considered correct if its IOU (intersection over union) with the true box is at least 0.8.}\label{table:eval}
\end{center}
\label{tab:figclsRes}
\end{table}



\section{Extracting Figures in Production}\label{deployment}


One of the primary applications for figure extraction is academic search. The system we described for figure extraction has been deployed at scale in the Semantic Scholar search engine\footnote{www.semanticscholar.org} to power the figures it serves along with other metadata from research papers. Currently, Semantic Scholar indexes more than 30 million research papers; it extracts and serves figures for 13 million of those papers, that is, the subset of papers which both allow figure extraction and have a PDF. Using a system built upon the techniques outlined in this paper, 96\% of the 13 million papers were successfully extracted allowing Semantic Scholar to index 4.5 figures (or tables) for each paper on average.


The deployed system distributes the figure extraction work across numerous machines using a job queue. Each machine runs multiple worker processes which perform the following tasks:

\begin{enumerate}
  \item Retrieve a unique identifier for a paper from the job queue.
  \item Pull the paper down from S3.\footnote{Simple Storage Service, AWS's highly scalable object store.}
  \item Run the DeepFigures library's figure extraction code locally to detect figures and extract captions.
  \item Crop the figures out of the rendered PDFs using the detected bounding boxes.
  \item Upload the detection results and cropped images into S3 for further processing.
\end{enumerate}

Once the cropped figures along with a JSON file describing the extraction results are in S3, they can be picked up by another data processing system such as Apache Spark for further transformations or indexing. In Semantic Scholar, this post-processing involves transforming the extraction results and indexing the figure metadata in Elasticsearch. From Elasticsearch, figures attached to a paper may be retrieved and their corresponding images served on the site directly from S3.


Due to the demanding computational requirements of the neural network models we use for figure detection, the use of GPU-capable machines can greatly accelerate the extraction process. The combination of I/O, network calls, rendering of PDFs, extracting captions from PDFs and neural network inference means that in order to fully utilize the GPU, CPU and network resources, it's helpful to run multiple processes extracting figures on one machine. In particular, the deployed system uses g2.xlarge instances in AWS EC2 with 8 worker processes to balance GPU and CPU requirements. Through AWS CloudFormation, the instance type is kept as a configuration option. The ability to configure the instance type, and other hardware decisions, enabled fast iteration to find a setup that properly balanced CPU, GPU and network resources. Such experimentation was crucial to finding machines with high throughput at a reasonable price.

Ultimately, the system is able to extract the figures from 15 papers per minute per machine and can scale horizontally. Running a single 12 page PDF through the whole pipeline, from queueing the paper through rendering and extraction, takes about 45 seconds. When the system is running at full speed, each paper takes around 30 seconds on average to be processed.

\begin{figure}
\includegraphics[width=0.99\linewidth]{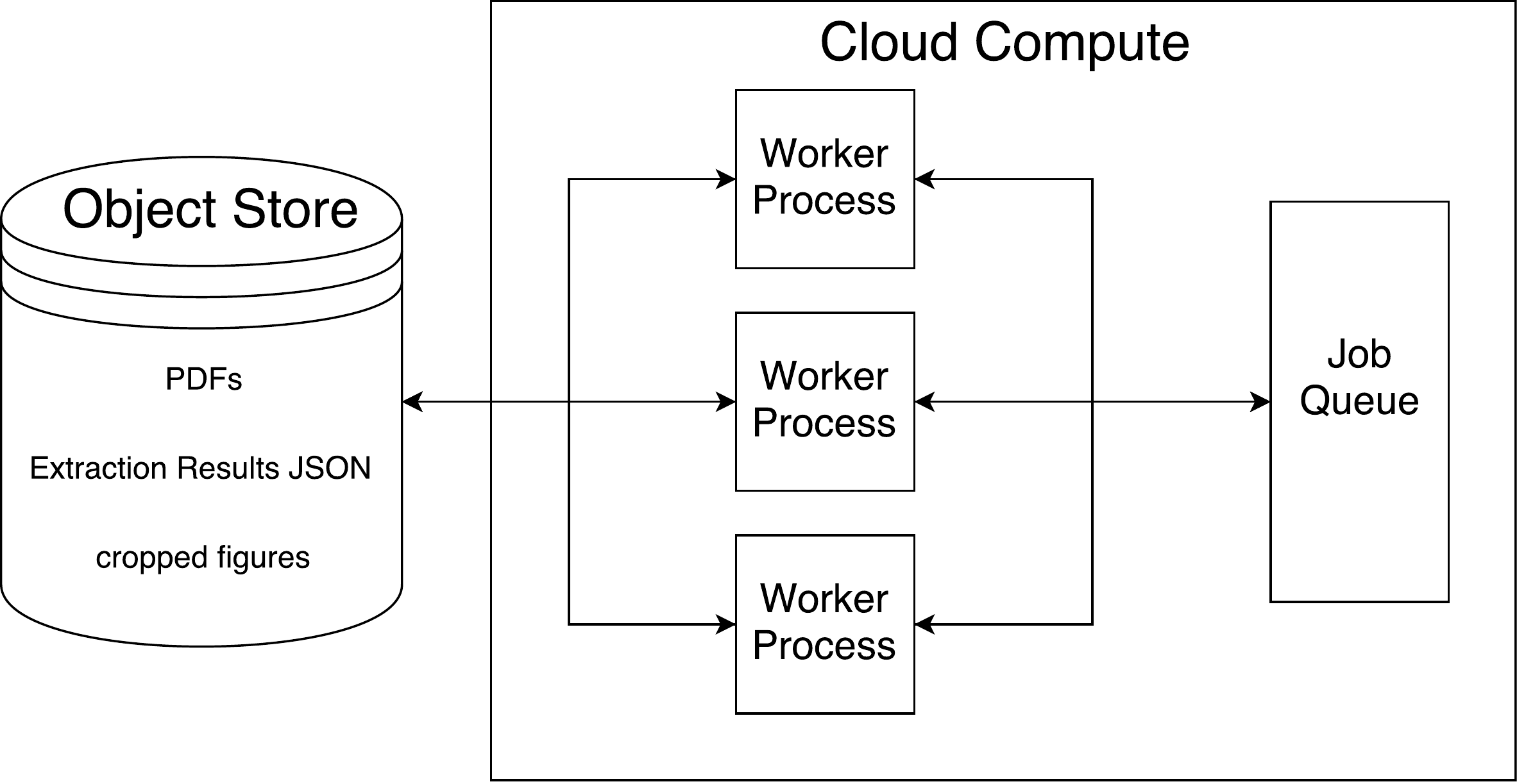}
\caption{Deployment architecture for the Deepfigures service.}
\label{fig:deployment-architecture}
\end{figure}
\section{Conclusion} 
In this work, we present a novel method for inducing high-quality labels for figure extraction in scientific documents.
Using this method, we contribute a dataset of 5.5 million induced labels with high accuracy, enabling researchers to develop more advanced methods for figure extraction in scientific documents.

We also introduce DeepFigures, a neural model for figure extraction trained on our induced dataset. DeepFigures has been successfully used to extract figures in 13 million papers in a large-scale academic search engine, demonstrating its scalability and robustness across a variety of domains.

Future work includes training a model to perform the full task of figure extraction end-to-end, including detecting and matching captions. This task could be aided by providing the network with additional information available from the PDF other than the rendered image, e.g. the locations of text and image elements on the page. Additionally, our data generation approaches could be extended to other information about papers such as title, authors, and sections; the distinctive visual characteristics of these elements as they appear in papers suggests neural detection models could be potentially useful.

\bibliographystyle{ACM-Reference-Format}
\balance
\bibliography{sigproc} 

\end{document}